\documentclass[aps,prl,twocolumn,superscriptaddress]{revtex4-1}
\usepackage{amssymb, amsmath,  hyperref, graphicx, bm, color}
\begin{document}
%------------------------------------------------------------------------------------------------------------------
%   macros
%------------------------------------------------------------------------------------------------------------------
\newcommand {\mb} {\mu_B}
\newcommand {\tpc} {T_\mathrm{pc}}
\newcommand {\tc} {T_c^0}
\newcommand {\ms} {m_s^\mathrm{phys}}
\newcommand {\ml} {m_l^\mathrm{phys}}
\newcommand {\g} {f_G}
\newcommand {\F} {F_\mathrm{reg}}
\newcommand {\zc} {z_c}
\newcommand {\todo}[1] {\textcolor{red}{#1}}
\renewcommand {\vec}[1] {{\mathbf #1 } }
%------------------------------------------------------------------------------------------------------------------
%   title
%------------------------------------------------------------------------------------------------------------------

%\begin{frontmatter}

\title{   
Probing Gluon Bose Correlations in Deep Inelastic Scatterings
}

%------------------------------------------------------------------------------------------------------------------
%     authors & affiliations
%------------------------------------------------------------------------------------------------------------------

\author{Alex~Kovner}
\affiliation{Physics Department, University of Connecticut, 2152 Hillside Road, Storrs, CT 06269, USA} 

\author{Ming~Li}
\affiliation{Department of Physics, North Carolina State University, Raleigh, NC 27695, USA}

\author{Vladimir~V.~Skokov}
\affiliation{Department of Physics, North Carolina State University, Raleigh, NC 27695, USA}
\affiliation{RIKEN/BNL Research Center, Brookhaven National Laboratory, Upton, NY 11973}

%------------------------------------------------------------------------------------------------------------------
%     abstract
%------------------------------------------------------------------------------------------------------------------

\begin{abstract} 
We study correlations originating from the quantum nature of gluons in a hadronic wave function. 
Bose-Einstein correlation between identical particles lead to the enhancement 
in the number of pairs of gluons with  the same quantum numbers and small relative momentum.
We show that these preexisting correlations can be probed in Deep Inelastic Scattering experiments at high energy. 
Specifically,  we consider diffractive dijet plus a third jet production. % in electron-hadron collisions at high energy. 
The azimuthal dependence  
displays a peak at the zero relative angle between the transverse momentum imbalance of the  photon-going dijet and 
the transverse  momentum of the hadron-going jet. 
Our calculations explicitly show  that the peak originates from Bose enhancement. 
Comparing electron-proton to electron-nucleus collisions, 
we demonstrate that the nuclear target 
enhances the relative strength of the peak. 
With the future high luminosity Electron-Ion Collider the proposed measurements of gluon Bose enhancement  
become experimentally feasible.
\end{abstract}
%------------------------------------------------------------------------------------------------------------------
%     keywords, PACS etc.
%------------------------------------------------------------------------------------------------------------------
%\begin{keyword}
%\end{keyword}
%\end{frontmatter}
%\pacs{pacs}
\date{\today}
\maketitle
%------------------------------------------------------------------------------------------------------------------
%     section: Introduction
%------------------------------------------------------------------------------------------------------------------
{\bf Introduction.}
The future Electron-Ion Collider will provide a unique opportunity to explore the multi-dimensional structure of protons and nuclei~\cite{AbdulKhalek:2021gbh,Aschenauer:2017jsk}.  
Although most of the experimental measurements and theoretical studies are focused on single-parton distributions \cite{Dumitru:2015gaa, Hatta:2016dxp, Dumitru:2018kuw, Mantysaari:2019hkq},  
complete theoretical and experimental understanding of a hadron wave function is not possible without observables sensitive to multi-parton correlations. The simplest objects that probe such correlations 
are multi-parton  distribution functions~\cite{Gaunt:2009re,Blok:2010ge, Diehl:2011yj, Blok:2013bpa,Diehl:2017wew} and generalized parton distributions~\cite{Diehl:2003ny,Boffi:2007yc}. 
Among these, gluon distributions play the most important role in high energy collisions. Indeed, 
the measurements at HERA established that 
the small x~\footnote{The Bjorken $x$ is the longitudinal momentum fraction of the hadron carried by a gluon.} tail of the hadronic wave function is dominated by gluons~\cite{Adloff:1999ah,Adloff:2000qk,Adloff:2001rw,Chekanov:2001qu}.
Thus it is imperative to identify observables sensitive to multi-gluon correlations.
The universal source of correlations between (identical) gluons is Bose enhancement due to quantum statistics.  In the hadron wave function this induces  the enhancement 
in the number of pairs of gluons with the same quantum numbers and small relative momentum~\cite{Dumitru:2010iy,Altinoluk:2015uaa,Kovner:2018azs}. These Bose-Einstein correlations have been suggested as a possible mechanism for producing ridge correlations in p-p scattering~\cite{Altinoluk:2015uaa,Altinoluk:2020psk}, however strong final state interactions make their effect difficult to isolate.
Here we show that these correlations can be probed in a clean straightforward way in Deep Inelastic Scattering (DIS) experiments. We also observe that the effect due to these correlations is {\it enhanced} by saturation effects in the target hadron.

\begin{figure}[!t]
    \centering
    \includegraphics[width=0.35\textwidth]{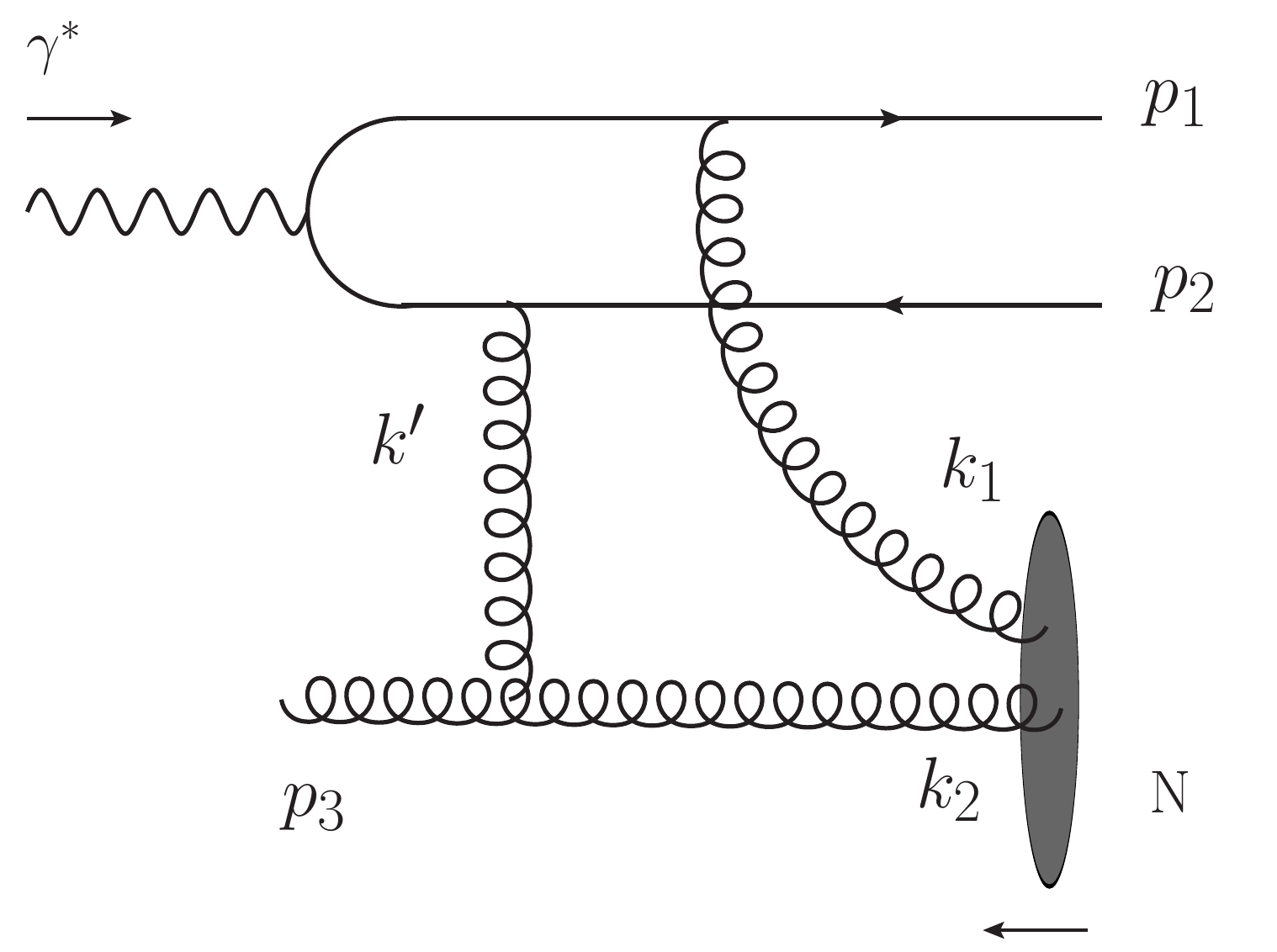}
    \caption{
		Schematic diagram showing the trijet production in $\gamma^{\ast}N$ collisions. Bose-Einstein correlations in the hadron wave function  
		lead to the increase in the cross-section of the trijet production,  when the transverse momenta $\vec{p}_{3} \approx  \pm (\mathbf{p}_{1}+\mathbf{p}_{2})$.  
		\label{fig:diag}
	}
\end{figure}

The general idea is as follows. 
At high energy, 
the intuitive picture of DIS in the infinite momentum frame is that of 
 the virtual photon %emitted from the electron in DIS 
 fluctuating into quark-antiquark pair (dipole) which scatters on the gluon field of the fast moving hadron target. 
As a result two  jets with the 
transverse momenta $\vec{p}_{1}$ and  $\vec{p}_{2}$ are produced. 
The transverse  momentum imbalance $\vec{\Delta} = \vec{p}_{1} + \vec{p}_{2} $  is acquired due to the interaction of the dipole with the hadron.
Consider now a final state which, in addition to the $q\bar q$ dijet, contains a gluon jet	  with transverse momentum $\vec{p}_{3}$ originating from the hadron.
In the hadronic wave function prior to the scattering this gluon is Bose correlated 
with an identical gluon (the two have momenta $\vec{k}_{1}  \approx \vec{k}_{2}  $). The exchange of the gluon  $\vec{k}_{1} $ between the hadron and the dijet leads to non-zero momentum imbalance  $|\vec{\Delta}| = |\vec{k}_{1}|$ when $\mathbf{k}' \approx 0$.   
In this situation, the momentum of the produced gluon does not change significantly ($\vec{p}_{3}  \approx \vec{k}_{2} $) and the primordial Bose-Einstein correlations should lead to the increase in the cross-section of the trijet production, 
when $\vec{p}_{3} \approx  \vec{\Delta} $. Note that one expects 
  the peak at $\vec{p}_{3} \approx  - \vec{\Delta} $ on the basis of the trivial physics of transverse momentum conservation, however the 
  peak  at  $\vec{p}_{3} \approx  \vec{\Delta}$ is {\it an evocative feature of Bose-Einstein enhancement}, see the diagram in Fig.~\ref{fig:diag}.

As we will see, the phase space where $\vec{k}'$ is small, is not  large. %To produce a diffractive di jet state, the third gluon must itself scatter off the dipole. 
This dilutes somewhat the correlations due to the  Bose enhancement in the hadron wave function. Our calculations however demonstrate that in a significant kinematical window
%although such dilution is present, 
the Bose-Einstein induced correlation is clearly seen in the trijet momentum distribution and may be sufficiently strong to be experimentally measurable. Moreover, we argue that  the unique features of this correlation facilitate its experimental identification and separation from the background. To minimize the effects due to Sudakov radiation~\cite{Sudakov:1954sw,Mueller:2013wwa} we  consider here a trijet configuration where the $q\bar q$ dijet is in the color singlet state. This still leaves room for Sudakov radiation from the gluon $\vec{p}_3$, however if the transverse momentum is not too large we don't expect this to qualitatively change the picture \footnote{The feasibility of measuring the dijets at the EIC was studied in \cite{Dumitru:2018kuw}. They might require the optimization of the radius of the jet cone.}. \\

{\bf Trijet production in high energy DIS.}
We focus on final states containing a color singlet $q\bar q$ dijet and the third jet originating from the nucleus, i.e. there is a large rapidity gap between the dijet and the third jet  (as opposed to Refs.~\cite{Iancu:2021rup, Boussarie:2016ogo} where the rapidity gap is between proton and the trijet).
We work in a frame where both the virtual photon and the hadronic target carry zero initial transverse
momentum, the photon carries large $p^+$ component of light cone momentum, while the hadronic target --  large $p^-$ component.
The virtual photon fluctuates into a dipole (quark-antiquark pair), which then eikonally  interacts
with the hadron producing the dijet. The final state gluon with momentum $\vec{p}_3$ can be thought as either originating from the hadronic target with momentum $\vec{k}_2$ and eikonally scattering on the dipole, or as originating from the dipole fluctuation with momentum $\vec{k}'$ and subsequently scattering on the hadron. It can be shown that both pictures mathematically lead to the same result \cite{KLS:2022}, and in this paper we adopt the former view.
The virtual photon-nucleon/nucleus cross section can be expressed as dipole-nucleon/nucleus cross section weighted over the dipole wave function. 
To lowest perturbative order this scattering  is given by a two-gluon process, see Fig.~\ref{fig:diag}.  One target gluon is absorbed by the dipole while
the other gluon interacts with the dipole and then emerges in the final state. 
Note that we are interested in different kinematics compared to commonly considered in the literature~\cite{Ayala:2016lhd,Boussarie:2016ogo,Iancu:2018hwa,Altinoluk:2020qet}, where all three jets are measured on the photon side.  
 Our calculation is performed in the  framework of  the Color Glass Condensate (CGC) effective theory~\cite{Iancu:2002xk,Gelis:2010nm,Kovchegov:2012mbw}.

The observable of interest is then given by
%\begin{widetext}
\begin{equation}
\begin{split}
&\frac{ d^3 N } {d^3 p_1  d^3 p_2  d^3 p_3 }
  = \sum_{\alpha_1,\alpha_2,\alpha_3}
\langle
\psi_F | \hat{d}^{\dagger}_{\alpha_1}(p_1)\hat{d}_{\alpha_1}(p_1)\\
&\qquad \times \hat{b}^{\dagger}_{\alpha_2}(p_2)\hat{b}_{\alpha_2}(p_2) \hat{a}^{\dagger}_{\alpha_3}(p_3)\hat{a}_{\alpha_3}(p_3) |\psi_F\rangle \, .\\
\end{split}
\end{equation} 
%\end{widetext}
Here $\alpha_{1,2,3}$ represent the spin (polarization) and color indices for the quark, antiquark and gluons, repsectively. The momenta are $p_{1,2}=(p^{+}_{1,2}, \mathbf{p}_{1,2})$ and $p_3=(p_3^-, \mathbf{p}_3)$. The expectation value of  quark, anti-quark, and gluon ($\hat{d}$, $\hat{b}$, and  $\hat{a}$)  number operators 
are evaluated over the final state 
\begin{equation}\label{psif}
|\psi_F\rangle = \hat{C}^{\dagger}\hat{S}   |\gamma^{\ast}\rangle \otimes |N\rangle. 
\end{equation}
Following the CGC framework, %effective theory, 
the initial hadron state $|N\rangle$  is represented in terms of  the state of the valence degrees of freedom  $|v\rangle$ and the vacuum of the soft gluons
in the presence of the valence sources $|s\rangle = \hat{C}_G|0\rangle$ as $|N\rangle=|v\rangle\otimes|s\rangle$,
with the coherent state operator \cite{Kovner:2005pe,Kovner:2005nq}
\begin{equation*}
\begin{split}
\hat{C}_G
=&\mathrm{exp} \bigg\{ i\int d^2\mathbf{x} \,b_i^a(\mathbf{x}) \int_{\Lambda^{-}e^{\Delta y}}^{\Lambda^-}\frac{dk^-}{\sqrt{2\pi} |k^-|} \Big(\hat{a}_i^{a\dagger}(k^-, \mathbf{x})\\
&\qquad  + \hat{a}_i^a(k^-, \mathbf{x})) \bigg\}\,.
\end{split}
\end{equation*}
Here $b_i^a(\mathbf{x})$ is the classical Weizs$\mathrm{\ddot{a}}$cker-Williams (WW) field generated by the valence degrees of freedom~\cite{Kovchegov:1997pc}.  The initial virtual photon state can be approximated by the quark-antiquark pairs $|\gamma^{\ast} \rangle \simeq\sum_{q\bar{q}} \Psi_{q\bar{q}}|q\bar{q}\rangle$ with $\Psi_{q\bar{q}}$ the dipole wavefunction whose explicit expression will be given below (see also  \cite{Beuf:2016wdz, Beuf:2017bpd}).%in later section 

To arrive at Eq.~\eqref{psif}, two effects are taken into account. {\it The first  effect} is due to eikonal $S$-matrix interaction between the dipole and the hadron:
\begin{equation}
\hat{S}=  \mathrm{exp}\left\{i\int d^2\mathbf{x} \,\hat{j}_{D}(\mathbf{x}) \frac{\partial^i}{\partial^2}\hat{A}^i( \mathbf{x})\right\}\,.
\end{equation}
Here the  color current operator of the dipole is
\begin{equation*}
\begin{split}
\hat{j}_D^a(\mathbf{y}) 
&=  g\sum_{s}  \int_{0}^{\infty} \frac{dk^+}{2k^+(2\pi)} \Big[ \hat{b}^{\dagger}_{h_1, s}(k^+,\mathbf{y}) t^a_{h_1h_2} \hat{b}_{h_2, s}(k^+,\mathbf{y}) \\
&\qquad + \hat{d}_{h_1, s}(k^+,\mathbf{y}) t^a_{h_1h_2} \hat{d}^{\dagger}_{h_2, s}(k^+,\mathbf{y}) \Big]\, 
\end{split}
\end{equation*}
and the gluon field operator $\hat{A}^i(\mathbf{x})$ has the conventional mode expansion in terms of gluon creation and annihilation operators in the light-cone gauge \cite{Brodsky:1997de, Collins:2011zzd}.
The hadron gluon field can be separated into the low and high longitudinal momentum modes.  
The latter can be treated as a classical field (the WW field), 
while the former, the quantum part,  in addition to the field produced by valence partons, also includes the field emitted by 
the higher longitudinal momentum gluonic modes~\cite{Baier:2005dv}  
$
\hat{A}_{i}(\mathbf{x}) = b_{i}(\mathbf{x}) + \delta \hat{A}_{i}(\mathbf{x})$.
At the  leading  order in the coupling constant,  
$
\delta \hat{A}_i(\mathbf{x}) \simeq \frac{\partial_i}{\partial^2} \hat{j}_G(\mathbf{x})$
with the gluon density operator  
\begin{equation*}
\hat{j}^a_G(\mathbf{x}) = ig f^{abc} \int_{k^- <\Lambda^-} \frac{dk^-}{2k^- (2\pi)} \hat{a}_i^{\dagger b}(k^-, \mathbf{x}) \hat{a}_i^c(k^-, \mathbf{x}).
\end{equation*}
The calculation is simplest in the dilute limit  and here 
 we 
expand the $S$-matrix operator to second order in the eikonal coupling.  This corresponds to two gluon exchange  in the amplitude. One of the four possible two gluon exchange digrams  is shown in Fig.~\ref{fig:diag}.

{\it The second effect} is due to final state radiations. 
It is accounted for by the ``dressing operator''  $\hat{C}^{\dagger}$. The reason this additional factor is necessary is that 
the eikonal  $S$-matrix propagates the projectile through the hadron, but does not include the evolution of the partons after they have left the interaction region.
This evolution results in emission of additional gluons which ``dress'' the bare partons by their WW fields as well as in ``recombination'' of some of the  outgoing soft gluons into the WW field of the outgoing fast partons. 
These processes happen in principle both on the hadron side (where we observe a single gluon jet) and on the dipole side (the quark-antiquark dijet). 
 However consistently with the lowest order approximation for the $\gamma^*$ wave function by a ``bare'' dipole state we neglect the additional emission effects form the dipole. 
On the other hand, the hadron side evolution cannot be neglected, since it directly affects the third jet. Taking it into account leads to the appearance of the dressing operator, so that our observable does not measure ``bare'' gluons,
but rather gluons that are not part of the WW field of the receding remnants of the hadron~\cite{Baier:2005dv}. 
We note that this is the high energy counterpart of the Faddeev-Kulish construction~\cite{Kulish:1970ut,Strominger:2017zoo} of dressed states in theories with massless gauge bosons. We thus simply have $\hat{C}^{\dagger} \approx \hat{C}^{\dagger}_G$.

\begin{figure}[!t]
    \centering
    \includegraphics[width=0.4\textwidth]{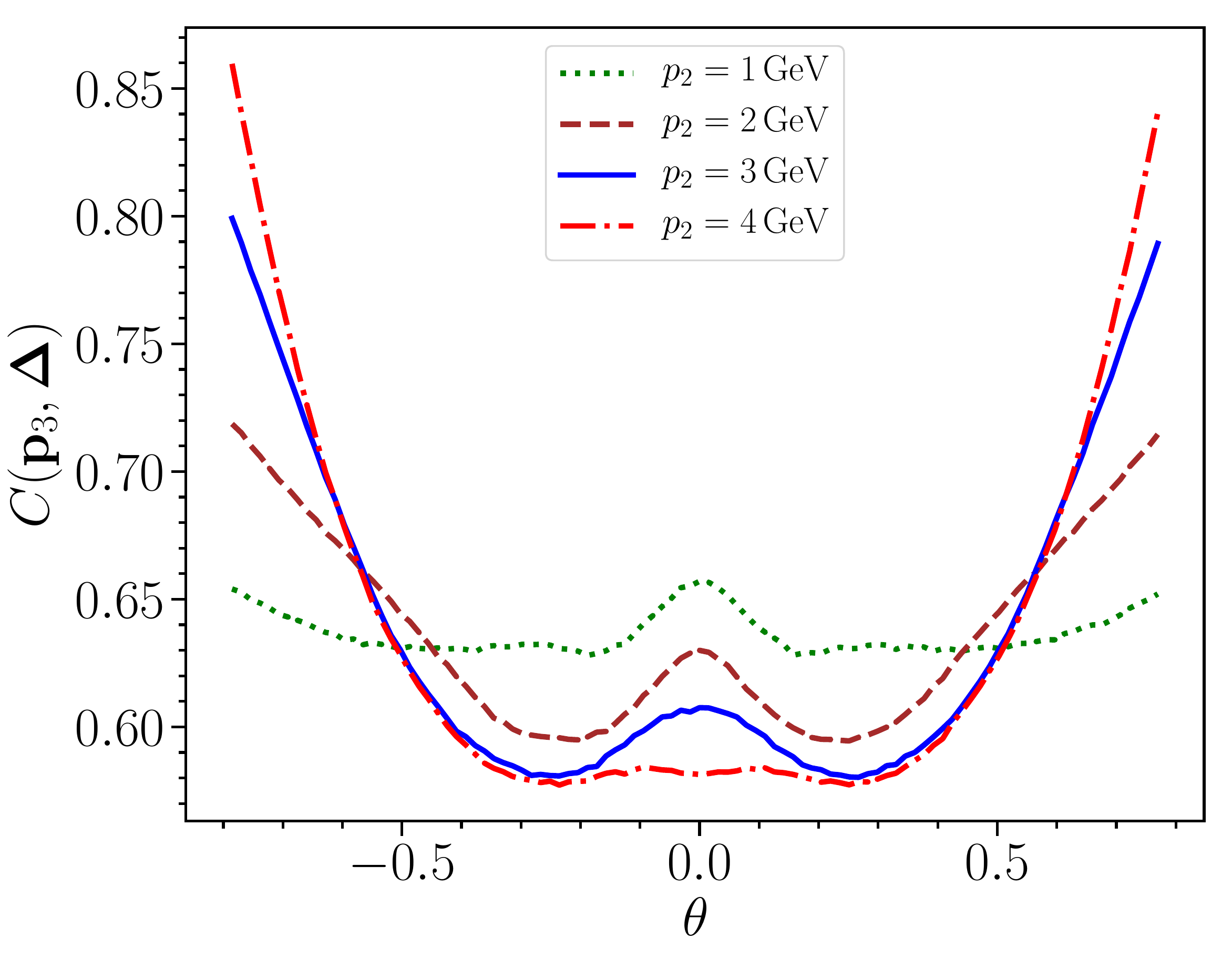}
    \caption{
		\label{fig:polar}
		The normalized trijet correlation as a function of the azimuthal angle between the vector of the momentum imbalance of the photon-going dijet $\vec{\Delta}$ and the momentum of the  nucleus going jet $\vec{p}_{3}$.   
		The magnitudes of both vectors are selected to be equal $|\vec{\Delta}| =  | \vec{p}_{3}|=10$ GeV and $p_1 = 10$ GeV, $Q_s=2$ GeV, $Q = 1$ GeV. } 
\end{figure}
\begin{figure}[!t]
    \centering
    \includegraphics[width=0.4\textwidth]{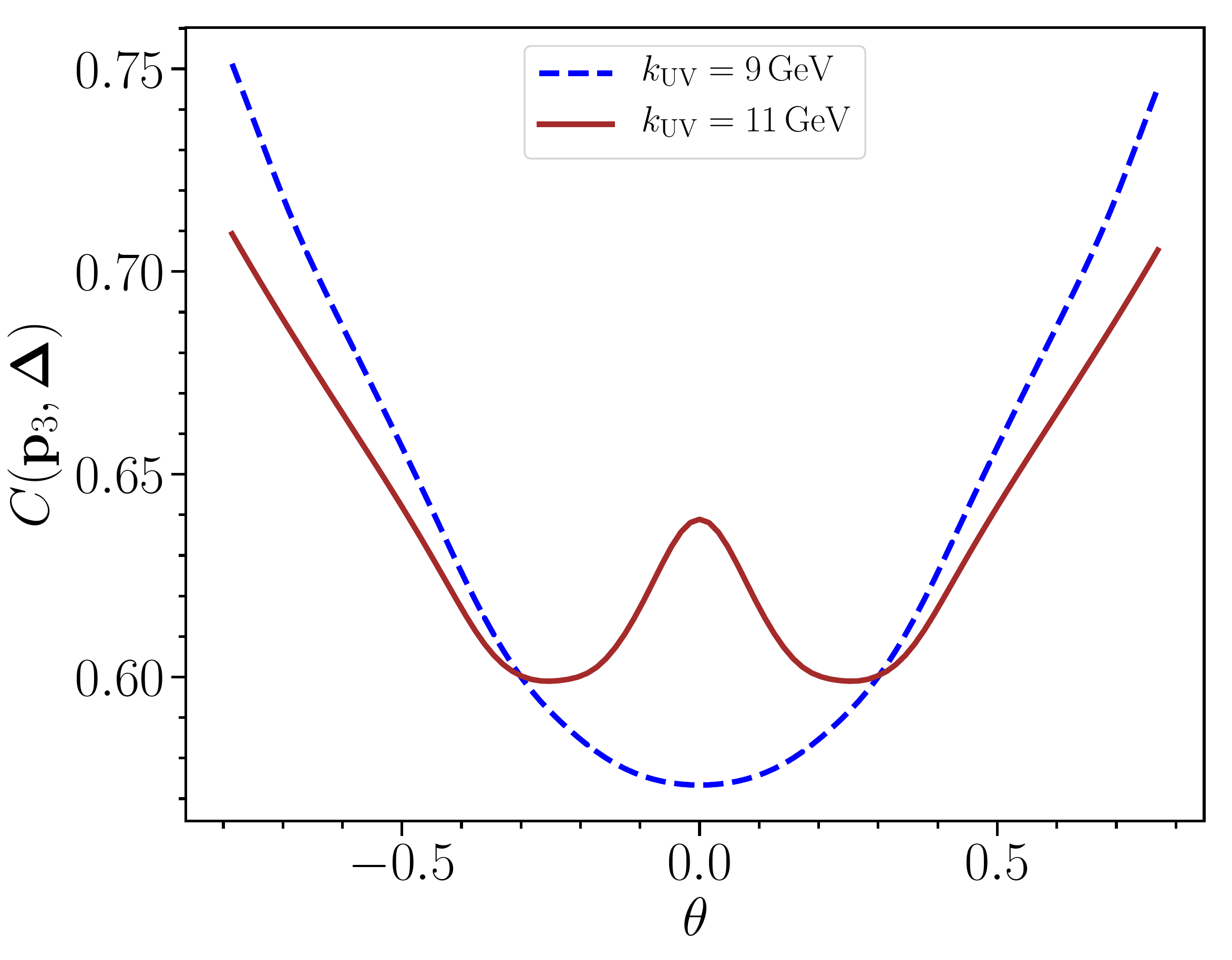}
    \caption{
		\label{fig:delta}
		The input parameters are the same as in  Fig.~\ref{fig:polar} with $p_2=2$ GeV, but varying the ultra-violet cutoff in the momentum integrals. 
		This illustrates that the peak at $\theta=0$ originates in the momentum region where $\mathbf{k} \approx -\Delta$ and thus is indeed due to the Bose-Einstein correlations.  
	}
\end{figure}

The trijet observable is then readily computed;  we present it in the following semi-factorizable form 
\begin{equation}\label{eq:trijet_factorization}
\begin{split}
\frac{ d^3 N } {d^3 p_1  d^3 p_2  d^3 p_3 }   
= &\int\frac{d^2\mathbf{k}}{(2\pi)^2}\frac{d^2\mathbf{l}}{(2\pi)^2} \, \mathcal{O}^{abcd}_{\mathrm{dipole}}(\{p_i\}; \mathbf{k},\mathbf{l}) \\
&\qquad\times \mathcal{O}^{abcd}_{\mathrm{hadron}}(\{p_i\}; \mathbf{k},\mathbf{l})\, .\\
\end{split}
\end{equation}

The part involving the hadron reads 
\begin{align}
\label{o}
%\begin{split}
\notag
&\mathcal{O}^{abcd}_{\mathrm{hadron}}(\{p_i\}; \mathbf{k},\mathbf{l}) = \frac{L_{j}(\mathbf{p}+\mathbf{l}, \mathbf{p}_3)}{l_{\perp}^2 |\mathbf{p}_1+\mathbf{p}_2+\mathbf{l}|^2}\frac{L_{j}(\mathbf{p}+\mathbf{k}, \mathbf{p}_3)}{k_{\perp}^2 |\mathbf{p}_1+\mathbf{p}_2+\mathbf{k}|^2}
\\ &\times
f^{bg h} f^{dg e} 
\langle v| \rho^c(\mathbf{l}) \rho^e(-\mathbf{p}-\mathbf{l}) \rho^a(-\mathbf{k}) \rho^h(\mathbf{p}+\mathbf{k})|v\rangle\, .
%\end{split}
\end{align}
Here  ${\mathbf p}  = \mathbf{p}_1+\mathbf{p}_2 + \mathbf{p}_3$, and  $L_{j}(\mathbf{p},\mathbf{k}) = (\frac{\mathbf{p}_j}{p^2}-\frac{\mathbf{k}_j}{k^2})$ is the Lipatov vertex. At the lowest order, the color charge density is linearly related to the WW field by $b_i^a(\mathbf{k}) = \frac{i\mathbf{k}_i}{\mathbf{k}^2} \rho^a(\mathbf{k})$. We regularize the infrared poles in $\mathcal{O}_{\mathrm{hadron}}$ by a non-perturbative scale of order $\Lambda_{\rm QCD}$. Our numerical results are insensitive to the exact value of this regulator. We varied it in the range $10^{-4}$ -$0.2\, \mathrm{GeV}$.  The four-density correlation function represents the \textit{two-particle transverse momentum dependent distribution} for gluons at small $x$, in which the Bose-Einstein correlation resides. 
To proceed, the statistical averaging over the color charge densities has to be performed.
From the  
 McLerran-Venugopalan (MV) model~\cite{McLerran:1993ka,McLerran:1993ni}  
\begin{equation}
 \langle v|  \rho^a(\mathbf{k}) \rho^b(\mathbf{l}) |v\rangle
 =  g^2 \mu^2 \delta^{ab}  \frac{ k^2}{k^2+Q_s^2}   \delta^{(2)}(\mathbf{k}+\mathbf{l}).
\end{equation}
The saturation momentum of the hadron ($Q_s\propto g^2 \mu$, see e.g.~\cite{Lappi:2007ku,Schenke:2012wb}), appears in this expression since color neutralization in a saturated hadronic system happens at this scale~\cite{Iancu:2002aq,McLerran:2015sva}.  
It naturally appears when higher order density corrections are included.

The averaging over the valence degrees of freedom in Eq.~\eqref{o} using the MV model  leads to  three different contractions. 
However the contraction between the color charge densities separately in the amplitude and in the complex amplitude 
vanishes for diffractive scattering due to color structure. The remaining two contractions are responsible for Bose-Einstein correlation \cite{Altinoluk:2015uaa} (see also \cite{Dumitru:2008wn}).

For the part involving the dipole, the terms 
contributing to the diffractive production are
\begin{equation}\label{eq:dipole_factor}
\begin{split}
&\mathcal{O}^{abcd}_{\mathrm{dipole}}(\{p_i\}; \mathbf{k},\mathbf{l})  = (2\pi)^3g^6\delta^{ab}\delta^{cd}\\
&\times\Big[\Psi^{\ast}_{r_1r_2}(p_1^+, \mathbf{p}_1; p_2^+, -\mathbf{p}_1) + \Psi^{\ast}_{r_1r_2}(p_1^+, -\mathbf{p}_2; p_2^+, \mathbf{p}_2) \\
&\qquad- 
(\mathbf{p}_1\to \mathbf{p}_1 + \mathbf{l}, \mathbf{p}_2\to \mathbf{p}_2 + \mathbf{l} )
\Big]\\
& \times \Big[ \left(\Psi_{r_1r_2}(p_1^+, \mathbf{p}_1; p_2^+, -\mathbf{p}_1)+\Psi_{r_1r_2}(p_1^+, -\mathbf{p}_2; p_2^+, \mathbf{p}_2)\right)\\
&\qquad 
- (\mathbf{p}_1\to \mathbf{p}_1 + \mathbf{k}, \mathbf{p}_2\to \mathbf{p}_2 + \mathbf{k} )
\Big]\,.
\end{split}
\end{equation}
The wavefunctions for the transversely and longitudinally polarized photons are  
\begin{equation}\label{eq:gammaT_WF}
\begin{split}
& %\delta(q^+-p_1^+-p_2^+) 
\Psi^{\rm T}_{r_1r_2}(p_1^+, \mathbf{p}_1; p_2^+, -\mathbf{p}_1) \\
=& -ee_f \delta_{r_1, -r_2}  
%\delta(1-z_1-z_2) 
\frac{2 (z_1-z_2+2\lambda r_1)\sqrt{z_1z_2}    \mathbf{p}_1 \cdot\mathbf{\epsilon}_{\lambda}}{\epsilon_f^2 + \mathbf{p}_1^2}\, , \\ 
%\end{split}
%\end{equation}
%while for the longitudinally polarized photon
%\begin{equation}\label{eq:gammaL_WF}
%\begin{split}
&% \delta(q^+-p_1^+-p_2^+) 
\Psi^{\rm L}_{r_1r_2}(p_1^+, \mathbf{p}_1; p_2^+, -\mathbf{p}_1) = -ee_f\delta_{r_1, -r_2} %\delta(1-z_1-z_2) 
\frac{ 4 z_1z_2 \epsilon_f}{\epsilon_f^2 + \mathbf{p}_1^2} .
\end{split}
\end{equation}
%We use  the common notation 
Here $\epsilon_f^2 = Q^2z_1z_2$, $z_{1,2} = p_{1,2}^+/q^+$,  note that $z_1 + z_2 =1$.  The photon polarization vector is $\epsilon_{\lambda=\pm 1}=(1,  \pm i)/\sqrt{2}$.% with $\lambda=\pm 1$.

Note that the observable in Eq.~\eqref{o} in our approximation does not depend on the rapidity of the gluon jet. This is always the case in high energy eikonal approximation, and we expect such a dependence to appear when the rapidity difference becomes too large~\cite{JalilianMarian:2004da,Kovner:2006wr}.

\begin{figure}[!t]
    \centering
    \includegraphics[width=0.4\textwidth]{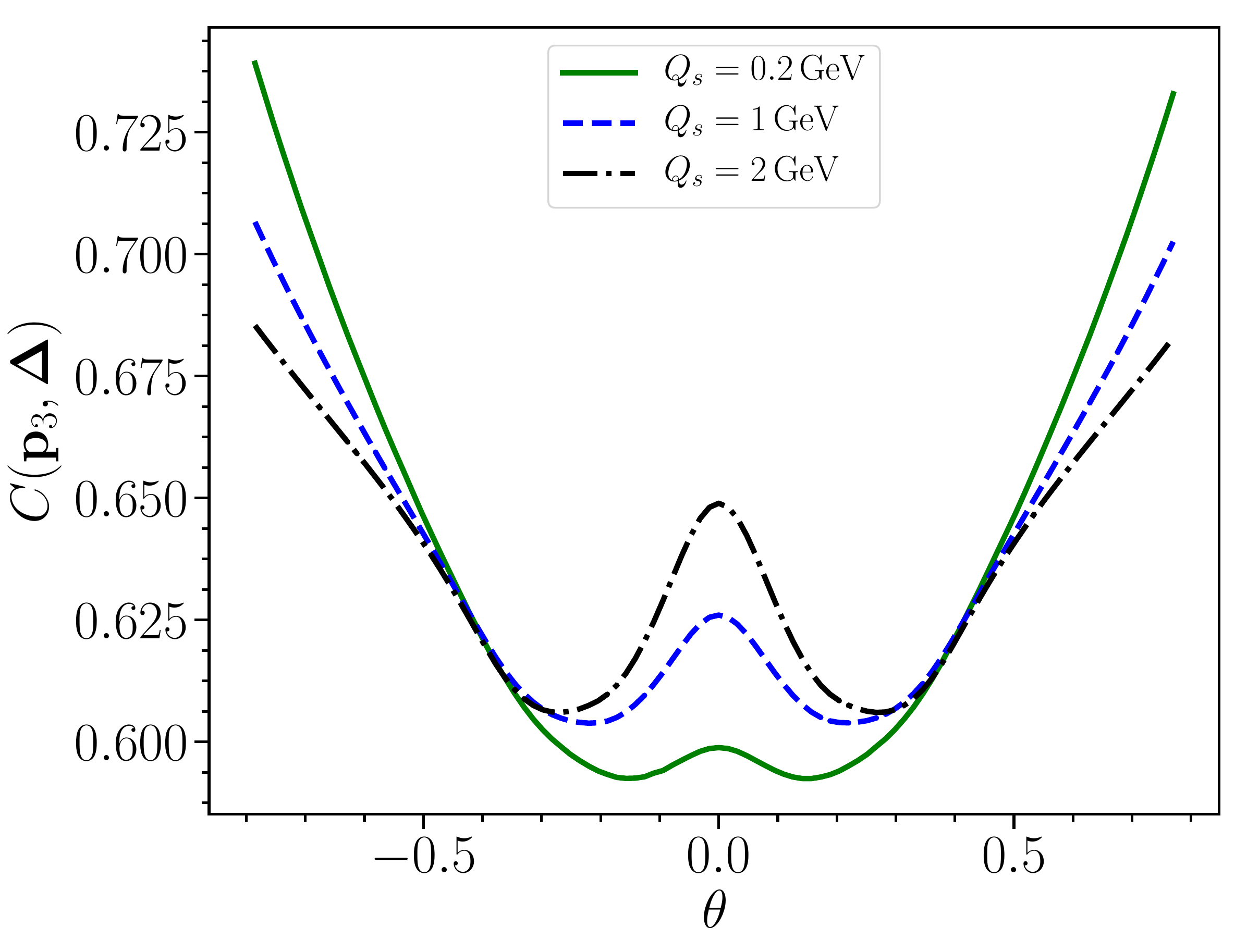}
    \caption{
		\label{fig:gammaPgammaA}
		The normalized tri-jet correlation for different values of the saturation scales. 
	}
\end{figure}

{\bf Numerical results and discussion.} 
 We are interested in the angular correlation and we numerically compute
the correlation as a function of $\theta$ ( with  $\cos{\theta } = \frac{\mathbf{p}_3\cdot\mathbf{\Delta}}{|\mathbf{p}_3||\mathbf{\Delta}|}$) given the magnitudes $|\mathbf{p}_3|=|\mathbf{\Delta}|$, 
\begin{align}
	\notag
	C_{\rm L,T}(p_3, \Delta, \theta; p_1, p_2) \equiv
	  \frac{1}{\mathcal{N}}\frac{ d^3 N _{\rm L,T}} {d^3 p_1  d^3 p_2  d^3 p_3 } ,  
	\notag
\end{align}
Here the normalization factor is computed by integrating over the angle $\mathcal{N} = \int d\theta d^3 N _{\rm L,T}/d^3 p_1  d^3 p_2  d^3 p_3$. We also need to specify the momentum magnitudes of the quark-antiquark jets $\mathbf{p}_1, \mathbf{p}_2$. In principle, one can integrate over all possible $\mathbf{p}_1,\mathbf{p}_2$ subject to the constraint $|\mathbf{\Delta}| =|\mathbf{p}_1+\mathbf{p}_2|$. However integrating over a large phase space masks the  Bose enhancement signal. Instead, we further select events  with particular values of $p_1, p_2$ for which the Bose enhancement signal is more prominent.  For illustrative purposes we choose  $\mathbf{p}_3 =(p_3, 0)$ along the $x$-axis and $z_1=z_2=1/2$ without loss of generality. We focus on transversely polarized virtual photon.\\

In Fig.~\ref{fig:polar}, we present numerical results for the dependence of   $C (p_3, \Delta, \theta)$ 
on the azimuthal angle $\theta$ with $p_3=\Delta = 10$ GeV at $p_1=10$ GeV for the transverse polarization of the virtual photon  with $Q=1$ GeV, and the nuclear target with the saturation momentum $Q_s=2$ GeV. The figure demonstrates a 
peak at zero azimuthal angle when $p_2\lesssim 4$ GeV. The zero-angle peak  at ${\mathbf p_3} = {\mathbf \Delta}$  is the salient feature of the Bose-Einstein correlation. The plots show the angular region $-\frac{\pi}{4}\leq \theta \leq \pi/4$. For angles close to $\theta =\pm \pi$ we observe, as expected a very large enhancement which arises due to the low momentum gluons in the target with $|{\mathbf k_1}|$ or $ |{\mathbf k_2}|$ $\approx \Lambda_{ \rm QCD}$, which is not related to Bose enhancement.

  We have also examined the values $p_1 = 11$-$15$ GeV (at fixed $p_3=\Delta$), and found that the zero-angle peak is also present although the range of the  relevant values of $p_2$ is smaller.  We have explored other momentum regions by  scaling  $p_3 =\Delta$ between 5 and 15 GeV, and have observed that the zero-angle peak persist.  Furthermore, we checked the Bose enhancement signals for longitudinal polarization of virtual photon. The enhancement signal here exists as well, albeit  it is less prominent compared than for the transverse polarization in the same kinematics.   

Note that the kinematics we explore is different from the frequently considered correlation limit $Q\gg \Delta$. In that regime, we analytically demonstrated the absence of  the zero-angle peak due to  inability of the exchanged gluon with momentum $<Q$ to resolve the structure of the dipole. The color neutrality then results in sizable  suppression through Eq.~\eqref{eq:dipole_factor}.

%The trijet production eq.~\eqref{eq:trijet_factorization} becomes a single momentum integration after ensemble averaging over the color charge densities. 
%In principle all regions of the phase space for a 2d momentum $\mathbf{k}$ contribute to the trijet production. 
Fig.~\ref{fig:delta} demonstrates that the zero-angle  peak in the correlation function is the direct reflection of the Bose-Einstein correlation in the hadronic wave function. Varying the UV cutoff we see that the peak comes from the momentum integration region $\mathbf{k} = -\Delta$. This corresponds to $\mathbf{k}' \approx 0$ in Fig. ~\ref{fig:diag}, which means $\mathbf{k}_2 \approx -\mathbf{p}_3$ and $\mathbf{k}_1 \equiv \mathbf{k}= -(\mathbf{p}_1+\mathbf{p}_2)$, thus in this region of phase space the trijet directly probes the momenta of the two gluons in the hadronic wavefunction.  %It is also worthy pointing out that when 
For $p_2\gtrsim 4$ GeV, the phase space region $\mathbf{k}\approx-\Delta$ is overwhelmed by contributions from the rest of the phase space $\mathbf{k}\gtrsim |\Delta|$; this leads to the suppression of the zero-angle peak  in Fig.~\ref{fig:polar} for  $p_2\gtrsim 4$ GeV.  

Finally, to utilize the unique characteristics of EIC to accelerate both protons and nuclei, we performed the calculations for different values of the gluon saturation scale $Q_s$. Fig.~\ref{fig:gammaPgammaA} demonstrates that Bose enhancement peak becomes more pronounced with increasing the gluon saturation scale.  Although the trijet momentum imbalance is mainly due to gluons with momentum larger than the gluon saturation scale, increasing the gluon saturation scale enhances the probability of larger momentum transfer between the dipole and the nuclear target. 
 Thus the correlation displays a clear dependence on the nuclear number and can serve as a sensitive probe of the saturation scale.    It is interesting that the main effect of saturation is to {\it enhance} the relative importance of the Bose-Einstein correlations in the trijet spectrum.

{\bf Conclusions.} We demonstrated that the measurement of gluonic Bose-Einstein correlations in hadronic wave function is accessible in DIS. The observable we propose is production of three jets at high energy with one of the jets separated in rapidity from the  other two (dijet) produced in the photon-going direction \footnote{For a configuration when $q\bar{q}$ dijet is not in a color singlet state,  the leading contribution would originate from single gluon exchange, which does not probe BE correlations and is not expected to contribute to near side correlations }. 
We showed the presence of zero angle correlation between the 
momentum imbalance of the dijet and the transverse momentum of the third jet. Its origin is due to preexisting Bose-Einstein correlations between gluons in the hadron wave function.
We demonstrated numerically the kinematic regions where this correlation 
becomes prominent. This zero-angle peak in angular correlation is a telltale feature of Bose-Einstein correlation.    We have also shown that the saturation in the hadronic wave function enhances the relative magnitude of the zero-angle peak  thereby providing a new sensitive probe of the saturation physics.\\

%------------------------------------------------------------------------------------------------------------------
%     acknowledgment
%------------------------------------------------------------------------------------------------------------------

{\bf Acknowledgments.}
M.L. and V.S.  are supported by the U.S. Department of Energy, Office
  of Science, Office of Nuclear Physics through the contract No. DE-SC0020081. 
A.K. is supported by the NSF Nuclear Theory grants 1614640 and 1913890.

  We thank  Haowu Duan and  Adrian Dumitru for illuminating discussions.

%------------------------------------------------------------------------------------------------------------------
%     bibliography
%------------------------------------------------------------------------------------------------------------------

  \bibliography{trijet}
%------------------------------------------------------------------------------------------------------------------
%     closing
%------------------------------------------------------------------------------------------------------------------
\end{document}